\documentclass[twocolumn,showkeys,aps,prb,showpacs]{revtex4-1}
\usepackage{graphicx}
\usepackage{amsmath}
\usepackage[CJKbookmarks,dvipdfm,colorlinks,linkcolor=blue,citecolor=blue]{hyperref}

\begin{document}

\title{Soft phonon modes driven huge  difference on lattice thermal conductivity  between topological semimetal WC and WN}

\author{San-Dong Guo$^1$}
\author{Peng Chen$^{2,3}$}
\affiliation{$^1$School of Physics, China University of Mining and
Technology, Xuzhou 221116, Jiangsu, China}
\affiliation{$^2$Beijing National Laboratory for Condensed Matter Physics,
Institute of Physics, Chinese Academy of Sciences, Beijing 100190, China}
\affiliation{$^3$School of Physical Sciences, University of Chinese Academy of Sciences, Beijing 100190, China}

\begin{abstract}
Topological semimetals are currently attracting increasing interest due to their potential
applications in topological qubits and low-power electronics, which are closely related to their thermal transport properties.
In this work, by solving the Boltzmann transport equation  based on first-principles calculations, we  systematically investigate
 the phonon transport properties of topological semimetal WC and WN.
The predicted room-temperature lattice thermal conductivities of WC (WN) along a and c directions are 1140.64  (7.47) $\mathrm{W m^{-1} K^{-1}}$ and  1214.69 (5.39)  $\mathrm{W m^{-1} K^{-1}}$. Considering the similar crystal structure
of WC and WN, it is quite interesting to find that the thermal conductivity of WC is more than two orders of magnitude higher than that of WN. It is  found that, different from WN,  the large   acoustic-optical (a-o) gap prohibits the acoustic+acoustic$\rightarrow$optical (aao) scattering,  which  gives  rise to very long phonon lifetimes, leading to ultrahigh  lattice thermal conductivity  in WC. For WN,  the lack of a-o gap  is due to soft phonon modes in optical branches,
which can provide more scattering channels for aao  scattering, producing very short phonon lifetimes. Further deep insight can be attained  from their different  electronic structures. Distinctly
different from that in WC, the density of states (DOS) of WN at the Fermi level  becomes very
sharp, which  leads to  destabilization of WN, producing  soft phonon modes. It is found  that the small shear modulus $G$ and $C_{44}$
limit  the stability of WN, compared with WC.
Our works provide  valuable information for  phonon transports in WC and WN, and motivate  further experimental works to study their lattice thermal conductivities.

\end{abstract}

\pacs{72.15.Jf, 71.20.-b, 71.70.Ej, 79.10.-n ~~~~~~~~~~~~~~~~~~~~~~~~~~~~~~~~~~~Email:guosd@cumt.edu.cn}

\maketitle

\section{Introduction}
Topological semimetals have attracted great research interest in condensed matter physics and material science\cite{q4,q5,q5-1,q7,q8,q11+0}.
In three dimensions, the band crossing of topological semimetals may occur at discrete points (Weyl/Dirac semimetals)\cite{q4,q5},  or along closed loops (topological nodal line semimetals)\cite{q8,q9}.
The $\mathrm{Na_3Bi}$  (Dirac semimetal) and TaAs (Weyl semimetal) have been confirmed by angle-resolved photoemission spectroscopy (ARPES)\cite{q5-1,q4,q10,q10-1,q10-2}.
Besides Dirac/Weyl fermion (four/two-fold degenerate point),
three-, six- or eight-fold band crossings have been proposed as new types of topological semimetals\cite{j1}.
The three-fold degenerate crossing points  have been predicted in a family of
two-element metals AB (A=Zr, Nb, Mo, Ta, W;
B=C, N, P, S, Te) with  WC-type structure \cite{q11-00,q11,q11-0},
 MoP of which  has been  experimentally confirmed  with triply degenerate nodal points (TDNPs), coexisting with the pairs of Weyl points\cite{q7}.

Topological semimetal  may have  substantial applications in topological qubits, spintronics and quantum computations\cite{k1}.
Efficient heat dissipation can ensure the reliability and stability of topological semimetal-based electronics devices. This is closely related to thermal transport, and high lattice thermal conductivity is beneficial to efficient heat dissipation\cite{k2}.
The lattice thermal conductivities of  topological semimetals TaAs, MoP, ZrTe and TaN have been predicted  by  first-principles calculations and the linearized phonon Boltzmann equation\cite{gsd1,gsd2,q12,q13,gsd3}.
The  lattice thermal conductivities of  TaAs, MoP and ZrTe are relatively low, and the room-temperature ones  are about 17$\sim$44 $\mathrm{W m^{-1} K^{-1}}$\cite{gsd1,gsd2,q12,q13}.  However, the  room-temperature lattice thermal conductivity of TaN  is ultrahigh,  838.62   $\mathrm{W m^{-1} K^{-1}}$ along a  axis and  1080.40   $\mathrm{W m^{-1} K^{-1}}$ along c axis, which is due to very large  a-o gap\cite{gsd3}.
In two-element metals AB, much different atomic masses of A and B atoms may lead to large a-o gap.
Based on mass difference factor,  the
WC and WN may have ultrahigh  lattice thermal conductivities\cite{gsd3}, which needs to be further confirmed by the first-principles calculations.
\begin{figure}
  \includegraphics[width=5cm]{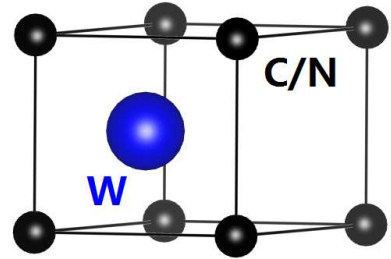}
  \caption{(Color online)The crystal structure of WC/N in one unit cell,  and the blue and black balls represent W and C/N atoms, respectively.}\label{st}
\end{figure}

In this work,  the phonon transport properties of WC and WN are studied  by solving the phonon Boltzmann transport equation.
It is found  that the thermal conductivity of WC is more than two orders of magnitude higher than that of WN.
This can be attributed to the very large a-o gap for WC, and the lack of a-o gap for WN  due to soft phonon modes in optical branches.
The soft phonon modes can be explained by DOS of WN at the Fermi level. Different from that in WC, the DOS of WN at the Fermi level  becomes very sharp, which  induces destabilization of WN, leading to soft phonon modes.
 Compared with WC and TaN,  the small shear modulus $G$ and $C_{44}$ may limit  the stability of WN.

The rest of the paper is organized as follows. In the next
section, we shall give our computational details about phonon transport. In the third section, we shall present phonon transports of WC and WN. Finally, we shall give our discussions and  conclusions in the fourth section.
\begin{figure}
  \includegraphics[width=8cm]{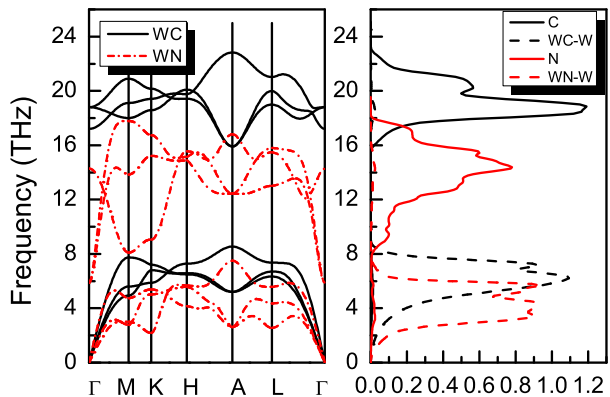}
  \caption{(Color online) Phonon dispersion curves of WC and WN with the
corresponding atom partial DOS. }\label{ph}
\end{figure}

\section{Computational detail}
We perform the first-principles calculations  within the projected augmented wave (PAW) method, as implemented in the VASP code\cite{pv1,pv2,pv3}.
With the plane-wave-cut-off energy of 500 eV, the generalized gradient approximation of the Perdew-Burke-Ernzerhof (GGA-PBE)\cite{pbe} is used  for the exchange-correlation functional. The 6s and 5d electrons of W, and 2s and 2p electrons of C/N  are treated as valance electrons.
The convergence criteria is with  energy convergences being less than  $10^{-8}$ eV.
The  lattice thermal conductivities of  WC and WN  are predicted  by solving linearized phonon Boltzmann equation with the single-mode relaxation time approximation (RTA),  as implemented in the Phono3py code\cite{pv4}. The lattice thermal conductivity can be attained by the following formula:
\begin{equation}\label{eq0}
    \kappa=\frac{1}{NV_0}\sum_\lambda \kappa_\lambda=\frac{1}{NV_0}\sum_\lambda C_\lambda \nu_\lambda \otimes \nu_\lambda \tau_\lambda
\end{equation}
where $\lambda$ is phonon mode, $N$ is the total number of q points sampling the  Brillouin zone (BZ), $V_0$ is the volume of a unit cell, and  $C_\lambda$,  $ \nu_\lambda$, $\tau_\lambda$   is the specific heat,  phonon velocity,  phonon lifetime.
The phonon lifetime $\tau_\lambda$ can be calculated by  phonon linewidth $2\Gamma_\lambda(\omega_\lambda)$ of the phonon mode
$\lambda$:
\begin{equation}\label{eq0}
    \tau_\lambda=\frac{1}{2\Gamma_\lambda(\omega_\lambda)}
\end{equation}
The $\Gamma_\lambda(\omega)$  takes the form analogous to the Fermi golden rule:
\begin{equation}
\begin{split}
   \Gamma_\lambda(\omega)=\frac{18\pi}{\hbar^2}\sum_{\lambda^{'}\lambda^{''}}|\Phi_{-\lambda\lambda^{'}\lambda^{''}}|^2
   [(f_\lambda^{'}+f_\lambda^{''}+1)\delta(\omega
    -\omega_\lambda^{'}-\\\omega_\lambda^{''})
   +(f_\lambda^{'}-f_\lambda^{''})[\delta(\omega
    +\omega_\lambda^{'}-\omega_\lambda^{''})-\delta(\omega
    -\omega_\lambda^{'}+\omega_\lambda^{''})]]
\end{split}
\end{equation}
in which $f_\lambda$ is the phonon equilibrium occupancy and
$\Phi_{-\lambda\lambda^{'}\lambda^{''}}$
is the strength of interaction among the three phonons $\lambda$, $\lambda^{'}$,
and $\lambda^{''}$ involved in the scattering.

Based on the supercell approach  with finite atomic displacement
of 0.03 $\mathrm{{\AA}}$,  the second- and third-order interatomic force constants (IFCs) can be attained.
For second-order harmonic IFCs, a 4 $\times$ 4 $\times$ 4  supercell  containing
128 atoms is used  with k-point meshes of 2 $\times$ 2 $\times$ 2.
According to the harmonic IFCs, phonon dispersions of WC and WN can be attained by  Phonopy package\cite{pv5},   determining  the allowed three-phonon scattering processes. The  group velocity  and specific heat can also  be compulated from phonon dispersion. For the third-order anharmonic IFCs,  a 3 $\times$ 3 $\times$ 3 supercells containing 54 atoms is used  with k-point meshes of 3 $\times$ 3 $\times$ 3.
 The three-phonon scattering rate can be attained by third-order anharmonic IFCs, and then  the phonon lifetimes can be calculated. To compute lattice thermal conductivities, the reciprocal spaces of the primitive cells  are sampled using the 20 $\times$ 20 $\times$ 20 meshes.
\begin{figure}
  \includegraphics[width=8cm]{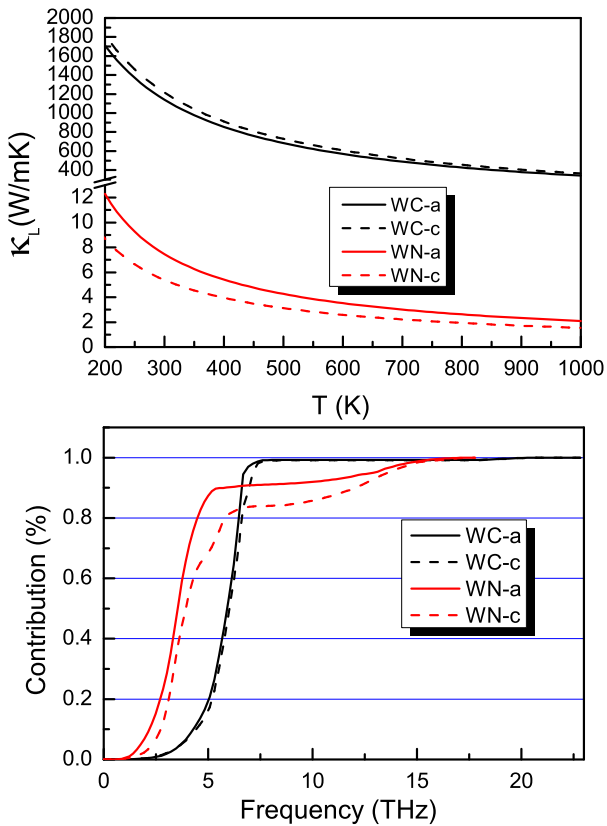}
  \caption{(Color online) The lattice thermal conductivities  of WC and WN  as a function of temperature, including a  and c directions; the cumulative lattice thermal conductivities (300 K) of  WC and WN divided by total lattice thermal conductivity with respect to phonon frequency, along a and c directions.}\label{kl}
\end{figure}

\begin{table*}
\centering \caption{The elastic constants ($C_{11}$, $C_{12}$, $C_{13}$, $C_{33}$, $C_{44}$ and $C_{66}$), bulk ($B$), shear ($G$) moduli  (in GPa) and $B$/$G$   of WC, WN and TaN. }\label{tab}
  \begin{tabular*}{0.96\textwidth}{@{\extracolsep{\fill}}cccccccccc}
  \hline\hline
 Name & $C_{11}$ & $C_{12}$& $C_{13}$&$C_{33}$ &$C_{44}$&$C_{66}$&$B$&$G$&$B$/$G$\\\hline\hline
WC&709.58&194.81&171.92&954.54  &300.76&257.38&380.87&291.72&1.31\\\hline
WN&637.37&200.34&283.65&756.06  &77.06&218.52&392.50&141.65&2.77\\\hline
TaN&566.40&128.23&62.41&706.39  &215.02&219.09&260.26&233.16&1.12\\\hline\hline
\end{tabular*}
\end{table*}

\section{MAIN CALCULATED RESULTS AND ANALYSIS}
The  low-temperature WC
and WN  have a hexagonal structure with space group  $P\bar{6}m2$ (No. 187), where W and C (N)
atoms occupy the 1d (1/3, 2/3, 1/2) and 1a (0,0,0) Wyckoff positions, respectively. 
The crystal structure is  shown in \autoref{st}.
The experimental lattice constants of WC ($a$=$b$=2.928  $\mathrm{{\AA}}$, $c$=2.835 $\mathrm{{\AA}}$ ) and WN ($a$=$b$=2.890  $\mathrm{{\AA}}$, $c$=2.830 $\mathrm{{\AA}}$ )\cite{q11-0}are used to investigate their  phonon transport properties.
The phonon dispersions of
 WC and WN along several high symmetry paths are plotted in \autoref{ph}, along with  atom partial DOS. The 3
acoustic and 3 optical phonon branches are observed due to two atoms in  one unit cell.
 There are no imaginary frequencies in the phonon dispersions of both WC and WN, which indicates the thermodynamic stability of WC and WN with hexagonal structure. A very wide a-o gap of 7.37 THz  is observed for WC, which is very close to  width of  acoustic branches (8.55 THz).
 The large   a-o gap  is  due to very different atomic masses of W and C atoms\cite{m1,m3}. Similar results can be found in topological semimetal TaN\cite{gsd3}.
  Compared to WC, the optical branches of WN exist soft phonon modes  near $\Gamma$, M and K points.
Then, the acoustic and optical
modes of  WN are not separated by the large band
gap, despite very different atomic masses  for W and
N atoms.  The size of a-o gap can produce very important effects on aao scattering\cite{h5,h6}.
It is found that acoustic (optical) branches of WC/N
are mainly contributed by the vibrations of
W (C/N) atoms  from atom partial DOS.
\begin{figure}
  \includegraphics[width=8cm]{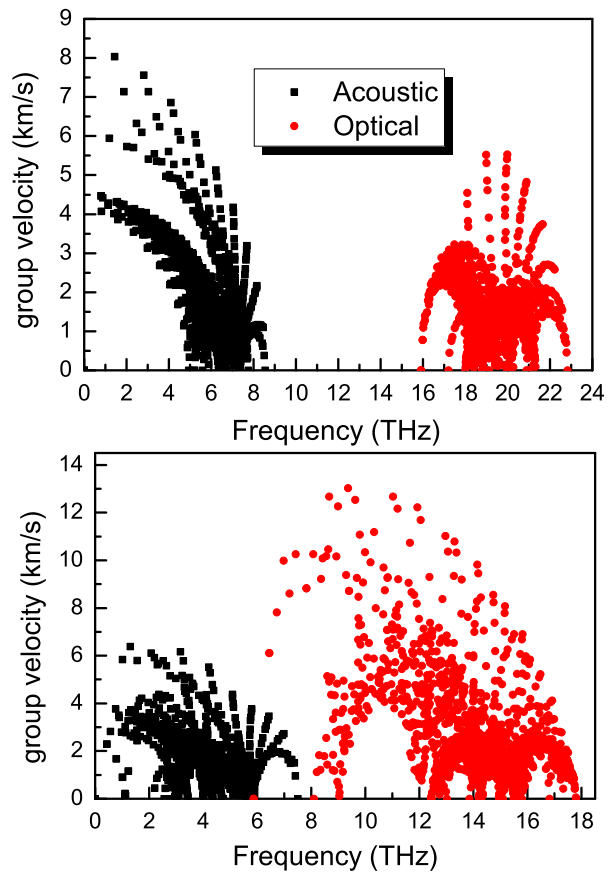}
  \caption{(Color online) The mode level phonon group velocities  of WC (Top) and WN (Bottom) in the first Brillouin zone. }\label{v}
\end{figure}
\begin{figure*}
  \includegraphics[width=14cm]{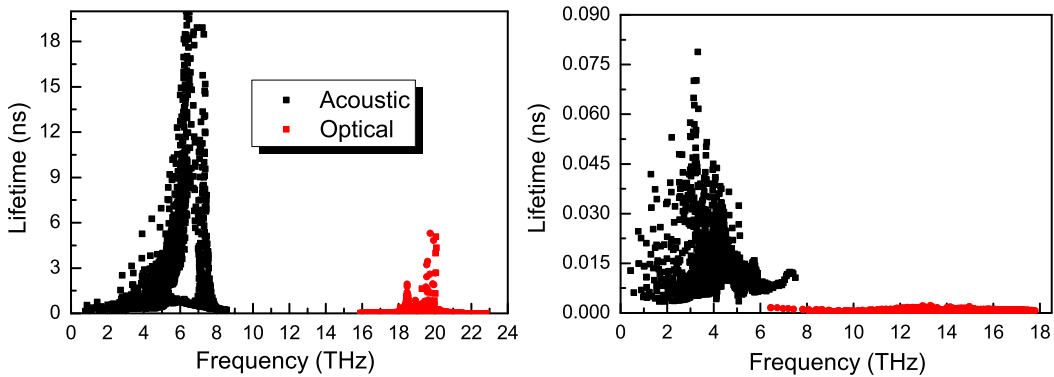}
  \caption{(Color online) The mode level   phonon lifetimes (300K)  of WC (Left) and WN (Right) in the first Brillouin zone.}\label{t}
\end{figure*}

The lattice thermal conductivities  of WC and WN along a and c directions as a function of temperature are shown in \autoref{kl}.
 Due to intrinsic enhancement of phonon-phonon scattering, the
lattice thermal conductivities  decrease  with increasing temperature in the considered temperature region.
The lattice thermal conductivity of WC
along  c direction  is higher than that along  a direction, which is consistent with that of TaAs, ZrTe, MoP and TaN\cite{gsd1,gsd2,q12,q13,gsd3}.
But it is opposite for WN. It is clearly seen that  the lattice thermal conductivity of WC is  hundreds of times higher than one of WN in the considered temperature region.
The room-temperature lattice thermal conductivities of WC along a and c directions are 1140.64  $\mathrm{W m^{-1} K^{-1}}$ and  1214.69  $\mathrm{W m^{-1} K^{-1}}$, and 7.47  $\mathrm{W m^{-1} K^{-1}}$ and  5.39  $\mathrm{W m^{-1} K^{-1}}$ for WN.
The lattice thermal conductivity  of WC is close to that of TaN\cite{gsd3}, which is very higher than that of TaAs, ZrTe and MoP\cite{gsd1,gsd2,q12,q13}.
However, the one of WN is lower than that of TaAs, ZrTe and MoP.
An anisotropy factor\cite{q12}  ($\eta=(\kappa_{L}(max)-\kappa_{L}(min))/\kappa_{L}(min)$)  is defined to measure the anisotropic strength of  lattice thermal conductivities  along a and c directions, and the corresponding value is 6.5\% for WC and 38.6\% for WN, implying weaker anisotropy for WC than WN. At room temperature, the cumulative lattice thermal conductivities divided by total lattice thermal conductivity (CLDT) with respect to frequency  along a and c directions are plotted  in \autoref{kl}. It is clearly seen  that acoustic branches dominate   lattice thermal conductivity for both WC and WN. For WC, the CLDTs along a and c directions almost coincide, showing weak anisotropy.    The CLDTs of WN along a and c directions show significant difference, which agrees well with strong anisotropy of lattice thermal conductivity. It is found that optical branches of WN have obvious contribution to total lattice thermal conductivity.

To gain more insight into huge difference on lattice thermal conductivity  between  WC and WN,  the mode level phonon group velocities
and lifetimes are shown in \autoref{v} and \autoref{t}, respectively.
  The most of group
velocities of acoustic branches  for WC are larger than those of WN, while  it is opposite for the optical branches.
Due to dominated contribution to total lattice thermal conductivity  from acoustic branches, group
velocities can partially explain higher lattice thermal conductivity for WC than WN. Because of group
velocities in the same order of magnitude between WC and WN, it can not account for huge difference on lattice thermal conductivity.
It  is found that  the most of group
velocities of WC for  acoustic branches are higher than those of optical branches, but it is opposite for WN.
From \autoref{t}, it is clearly seen that   phonon lifetimes of WC are   hundreds of times longer  than ones of WN, which leads to extremely higher lattice thermal conductivity for WC than WN. Such long phonon lifetimes can also be found in TaN\cite{gsd3}.
Although acoustic branches dominate heat transport, the  optical branches can  provide
 scattering channels for the acoustic modes to achieve three phonon scattering, especially for aao scattering.
 For WC, the huge a-o gap impedes  the annihilation process of two acoustic phonon
modes into one optical one due to the requirement on energy conservation, which leads to  weak phonon-phonon scattering
rate,  resulting in very long phonon lifetimes.
Similar mechanism can also be found in TaN\cite{gsd3}, BAs\cite{h5} and AlSb\cite{h6}.
For WN, the aao scattering can be easily achieved due to absent a-o gap, which gives to very short
phonon lifetimes.

\begin{figure}
  \includegraphics[width=8cm]{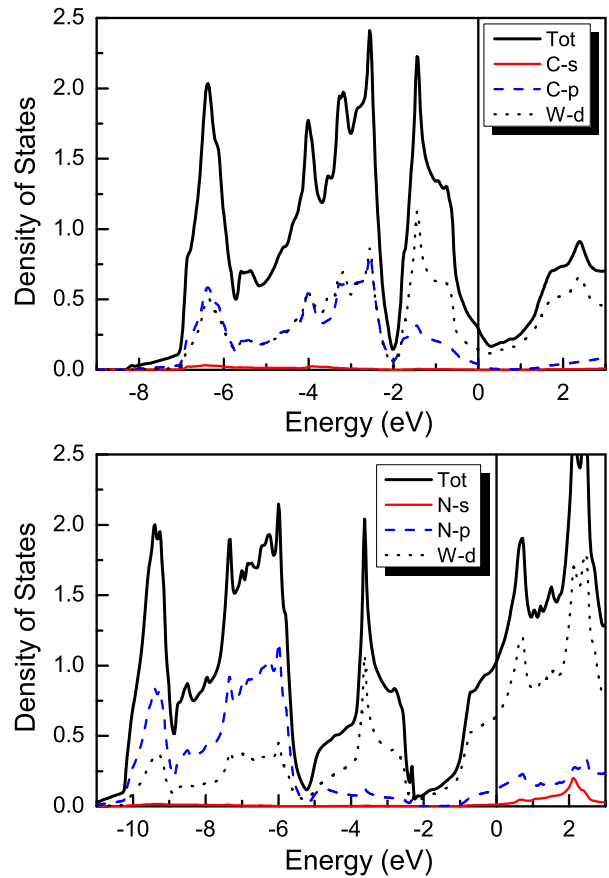}
  \caption{(Color online) The total and  projected DOSs of WC (Top) and WN (Bottom).}\label{dos}
\end{figure}
\begin{figure*}
  \includegraphics[width=12cm]{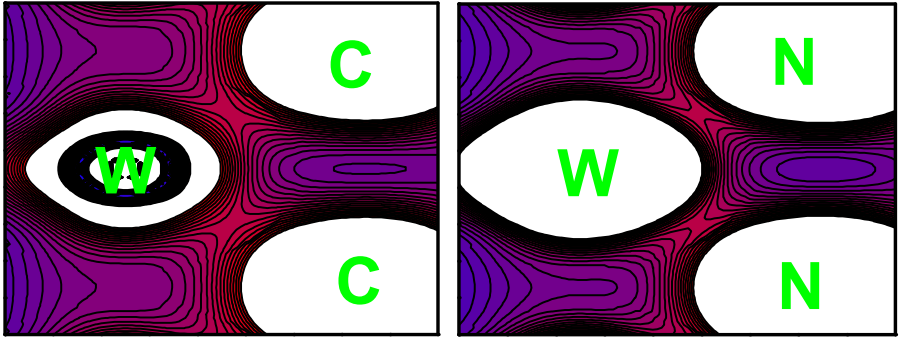}
  \caption{(Color online) The charge density distributions of WC (left) and WN (Right) [unit:$\mathrm{|e|}$/$\mathrm{bohr^3}$] in the (110) plane. The charge density value increases from blue to red.}\label{den}
\end{figure*}

In spite of  very different atomic masses  for W and
N atoms, the a-o gap of WN disappears, which is due to   soft phonon modes in optical branches.
 The soft phonon modes  mean that WN has weaker stability than WC.
 In order to get a deep insight into the  very different nature between WC and WN, we further perform analysis from the view of
electronic structures.   The atomic structure and  behavior of electrons fundamentally
determine all the properties of a material\cite{hm}. The total
and partial DOSs of WC and WN are plotted in \autoref{dos}. For WC, the strongly hybridized
W-5d and C-2p states are observed from -7.0 eV to -2.0 eV, which  is indicative of strong covalent
bonding.  However,  a weak  ionic component is observed due to a partial charge transfer from W to C atoms from \autoref{den}.
The  W-5d states have main contribution from -2.0 eV to Fermi level.
The Fermi level is  close to the DOS minimum, which qualitatively declares the stability of WC.
If the Fermi level lies at the local DOS maximum,  this will
induce destabilization. A similar hybridization picture is observed for WN, but an additional electron   further fills  W-5d states compared
with WC, where  the DOS at the Fermi level  increases
sharply.   This leads  to a destabilization of WN,  inducing soft phonon modes in optical branches.
Compared with C-2p and W-5d states of WC, the  hybridization of W-5d and N-2p states of WN decreases due to enhanced energy separation between N-2p
and W-5d states.  This leads to weaken  covalent  bonds compared with isostructural WC. The  weakness of covalent  bonds from WC to WN can also be easily observed from charge density distributions in \autoref{den}.

\begin{figure}
  \includegraphics[width=8cm]{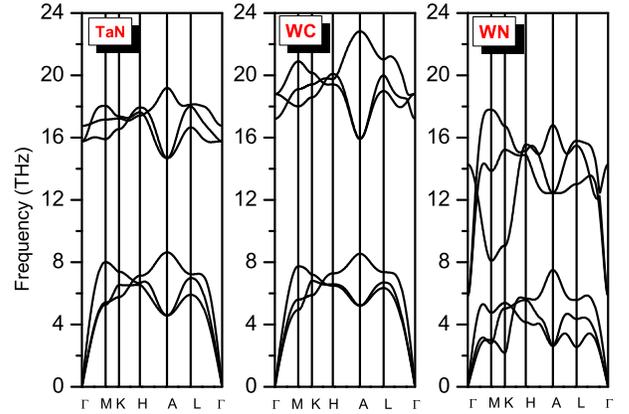}
  \caption{(Color online)Phonon  dispersion curves of TaN, WC and WN. }\label{ph-1}
\end{figure}

The weaker stability for WN than WC can also be understood by their elastic properties. Due to the hexagonal  structure for WC and WN,  five
independent elastic constants  ($C_{11}$, $C_{12}$, $C_{13}$, $C_{33}$ and $C_{44}$) are listed in \autoref{tab}, which are close to previous calculated values\cite{el1}. These elastic constants  satisfy
the well-known Born stability criteria\cite{el,q15}:
\begin{equation}\label{e1}
C_{44}>0,C_{11}>0
\end{equation}
\begin{equation}\label{e1}
 C_{11}>|C_{12}|
\end{equation}
\begin{equation}\label{e1}
(C_{11}+2C_{12})C_{33}>2C_{13}^2
\end{equation}
These mean  that WC and WN are mechanically stable.  By Voigt-Reuss-Hill
approximations,  the bulk and  shear  modulus of WC and WN can be obtained from elastic constants.
The Voigt's, Reuss's and Hill's bulk modulus can be calculated  by the following equations:
\begin{equation}\label{4}
    B_V=\frac{1}{9}(2C_{11}+C_{33}+2C_{12}+4C_{13})
\end{equation}
\begin{equation}\label{5}
    B_R=(2S_{11}+S_{33}+2S_{12}+4S_{13})^{-1}
\end{equation}
\begin{equation}\label{6}
    B_H=\frac{1}{2}(B_V+B_R)
\end{equation}
The Voigt's, Reuss's and Hill's shear modulus can be attained by using these formulas:
\begin{equation}\label{4}
    G_V=\frac{1}{15}(2C_{11}+C_{33}-C_{12}-2C_{13}+6C_{44}+3C_{66})
\end{equation}
\begin{equation}\label{5}
    G_R=[\frac{1}{15}(8S_{11}+4S_{33}-4S_{12}-8S_{13}+6S_{44}+3S_{66})]^{-1}
\end{equation}
\begin{equation}\label{6}
    G_H=\frac{1}{2}(G_V+G_R)
\end{equation}

The $S_{ij}$ are the elastic compliance constants. The Hill's bulk and  shear  modulus are listed in \autoref{tab}.
It is assumed that the hardness of materials can be measured by  bulk modulus or shear modulus\cite{h1}. For transition-metal carbonitrides, the magnitude of  $C_{44}$ may be a better hardness
predictor\cite{h2,h3}. Therefore, it may be concluded that WC is more harder than  WN.
The $B$/$G$ can  measure that a  material is  ductile ($B$/$G$$>$1.75) or brittle ($B$/$G$$<$1.75).
Based on $B/G$ in \autoref{tab}, for WC, the brittle character is dominant. For WN, the increase in the plasticity  may be  due to enhanced metallicity, caused by high DOS at the Fermi level.  It is also found that  the
shear modulus $G$ and $C_{44}$ may  limit the stability of WN.
\begin{figure}
  \includegraphics[width=8cm]{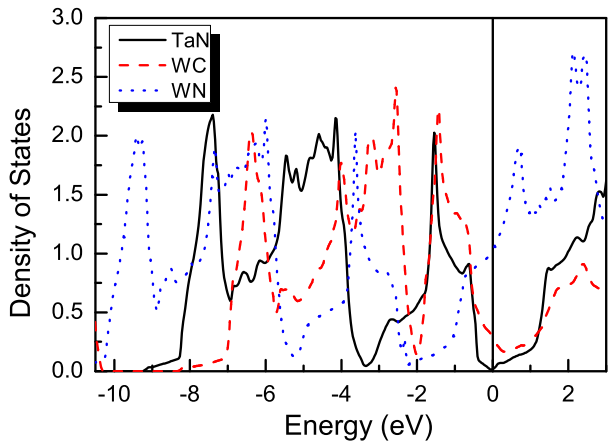}
  \caption{(Color online) The total  DOSs of TaN, WC and WN.}\label{dos-1}
\end{figure}

\section{Discussions and Conclusions}
The significant mass differences between the constituent atoms can induce a-o gap\cite{m1,m3}. In some  2D materials, gap formation
mechanism  is due to the violation of  reflection symmetry selection
rule in the harmonic approximation\cite{m5,m6}. Based on mass difference factor,  defined as $\delta=(M_{max}-M_{min})/M_{min}$,
 WC and WN may be potential candidates with ultrahigh  lattice thermal conductivity due to $\delta$ of WC and WN being  larger than 10\cite{gsd3}.
Calculated results show that WC indeed has ultrahigh  lattice thermal conductivity, which is the same with TaN with  $\delta$ being 11.92\cite{gsd3}.
However, the WN has very low lattice thermal conductivity. The phonon dispersions of TaN, WC and WN  are plotted in \autoref{ph-1}.
The TaN and WC have very similar outlines of phonon dispersion, and  a very large a-o gap for TaN (6.05 THz) and WC (7.37 THz) is observed.
The very large a-o gap  can  produce  inefficient aao scattering, which leads to long phonon lifetimes, giving rise to ultrahigh  lattice thermal conductivity.
A significant difference is observed between WC/TaN and WN, and a-o gap of WN  is absent.  As a result, the optical branches of WN can  provide
 more scattering channels for the acoustic modes, which leads to strong aao scattering,  giving rise to very low lattice thermal conductivity.

Compared with TaN and WC, the lack of a-o gap for WN is due to soft phonon modes in optical branches. The softness of phonon modes can be explained by  electronic structures. The DOSs of TaN, WC and WN  are plotted in \autoref{dos-1}. It is clearly seen that  the Fermi level for TaN and WC
is close to the DOS minimum, while the DOS at the Fermi level for WN  increases
sharply, which leads to destabilization of WN, inducing soft phonon modes.  In fact, they have similar electronic structures. The total number of valence electrons in TaN and WC is ten.
Adding one more  electron for WN  produces   softness of the phonon  modes due to further filling of W-5d states. Based on elastic properties, it is concluded  that the parameter $C_{44}$ and shear modulus $G$ may limit  the stability of WN. Based on our calculated results, the  mass difference factor and total number of valence electrons in  two-element metals AB  with  WC-type structure  determine whether they have ultrahigh  lattice thermal conductivity.

In summary,  we investigate the phonon transport properties of WC and WN by solving the Boltzmann
transport equation based on first-principles calculations.  Although the WC and WN  possess  similar crystal
structure,  they show very  diverse lattice thermal conductivity.
The room-temperature lattice thermal conductivity for  WC is  hundreds of times higher than one of WN.
The ultrahigh  lattice thermal conductivity in WC is due to
the extremely large a-o gap, which  strongly restricts aao scattering through
the energy conservation, producing very long phonon lifetimes.
The very low lattice thermal conductivity in WN is due to softness of phonon modes in optical branches, which can provide
 more scattering channels for aao  scattering, leading to short phonon lifetimes.
 We further understand the huge difference between WC and WN from the view of electronic structures, considering that all the properties are fundamentally determined by the electronic structures.  Distinctly
different from that in WC and TaN, the DOS of WN at the Fermi level  becomes very
sharp, which  produces  destabilization of WN, leading to soft phonon modes. It is also found that the elastic  parameter
limiting the stability of WN  is the
shear modulus $G$ and $C_{44}$.
Our works provide fundamental understanding of
phonon transport in WC and WN, which will enrich the studies of  phonon
transport in topological semimetals,  and will motivate farther experimental works.

\begin{acknowledgments}
This work is supported by the National Natural Science Foundation of China (Grant No.11404391). We are grateful to the Advanced Analysis and Computation Center of CUMT for the award of CPU hours to accomplish this work.
\end{acknowledgments}

\end{document}